\documentclass{ifacconf}

\usepackage{graphicx}      
\usepackage{natbib}        
\usepackage{amsmath}
\usepackage{url}

\usepackage{calc}
\newlength\myheight
\newlength\mydepth
\settototalheight\myheight{Xygp}
\usepackage{xcolor}

\usepackage{csquotes}
\begin{document}
\begin{frontmatter}

\title{LU-PZE: Lund University Pole-Zero Explorer\thanksref{footnoteinfo}}

\thanks[footnoteinfo]{This work was partially supported by the Wallenberg AI, Autonomous Systems and Software Program (WASP) funded by the Knut and Alice Wallenberg Foundation. All authors are members of the ELLIIT Strategic Research Area at Lund University.}

\author[First]{Pex Tufvesson} 
\author[Second]{Frida Heskebeck}

\address[First]{Ericsson Research, Lund, Sweden. Department of Automatic Control, Lund University, Lund, Sweden (e-mail: pex.tufvesson@ericsson.com).}
\address[Second]{Department of Automatic Control, Lund University, Lund, Sweden (e-mail: frida.heskebeck@control.lth.se).}

\begin{abstract}                
LU-PZE is an interactive tool for illustrating fundamental
concepts related to control theory, covering the relation
between transfer functions, pole-zero plots, step responses,
Bode plots, and Nyquist diagram. The tool gamifies education with dynamic assignments and quizzes. The tool is straightforward to use since it is web-based. \url{https://lu-pze.github.io}

\end{abstract}


\end{frontmatter}

\section{Introduction}\label{sec:introduction}
\begin{figure}
\begin{center}
\includegraphics[width=8.9cm]{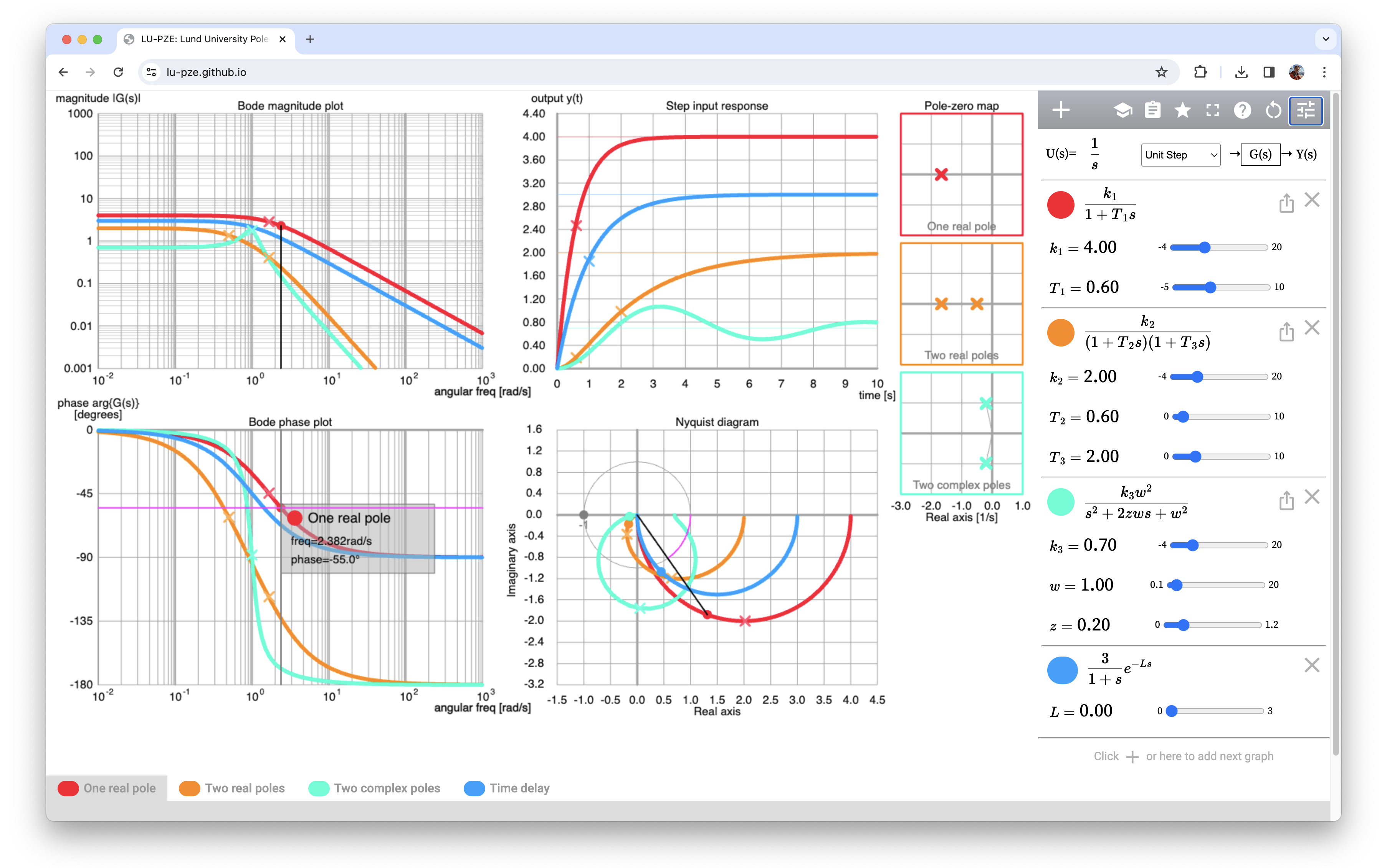}    
\caption{Default view of the LU-PZE interactive automatic control theory playground.} 
\label{fig:whole_screen}
\end{center}
\end{figure}

A student who has passed a basic course in automatic control should be able to design controllers to solve practical problems in the industry\footnote{\url{https://kurser.lth.se/lot/course-syllabus-en/23_24/FRTF05}}. This means the student needs a good knowledge of control theory and hands-on experience. One way to aid the students in gaining an understanding of the theoretical parts of automatic control is to use interactive tools in the education that illustrate these concepts. Using interactive tools to reinforce the concepts learned in a course is nothing new. However, this paper presents Lund University Pole-Zero Explorer (LU-PZE), as seen in Figure~\ref{fig:whole_screen}, an interactive tool run in the web browser that requires no software installation, as opposed to many other interactive tools, making it extraordinarily easy to use. Gamification of education has been shown to increase student motivation in higher education \citep{zeybek_gamification_2024,buckley_gamification_2016}. Therefore, LU-PZE also includes assignments and quizzes, which provide a gamified version of the education, challenging the student to discover the control theory concepts illustrated in the tool.

The purpose of all teaching is for the students to learn the course content. As described by, for example, Bloom's taxonomy \citep{bloom_taxonomy_1956,anderson_taxonomy_2000} or the SOLO-taxonomy \citep[p.~86-91]{biggs_teaching_2011}, students that have a high level of understanding of a topic can make connections between different aspects of the course and apply their knowledge in new ways. In control theory, this could concretely be formulated as the students at least knowing the relation between a transfer function, a step response, the poles and zeros, a Bode plot, and a Nyquist diagram\footnotemark[1]\citep{glad_control_2000}. If students understand the connections between these topics, they are moving in the right direction toward comprehending control theory and becoming proficient control engineers.

Research has shown that students engaging in active learning perform better on exams. As the name indicates, active learning means that the students actively work with the course content. The opposite is passive learning, which for example can occur while the students are listening during a lecture (\citeauthor{elmgren_academic_2018}, \citeyear{elmgren_academic_2018}, p.~66-67; \citeauthor{freeman_active_2014}, \citeyear{freeman_active_2014}). Many teaching activities engage students in active learning, for example, discussion seminars \citep{heskebeck_discussion_2023}, exercises, and lab sessions. In addition to discussions, calculations, and practical experiences, interactive tools are used in many control courses to engage students in active learning and aid their learning. The interactive tools could be a classroom response system (CRS) that are typically used to engage students during lectures \citep{garcia-lopez_implications_2022} or an online assessment tool. In this paper, the term interactive tool refers to a tool that illustrates some concept from a course that the students can explore for themselves which helps the student learn the course better.

Computer tools have been used to analyze control systems for almost as long as personal computers have been around \citep{astrom_computer_1983}. Many interactive tools, with the purpose of aiding students in their learning, have been developed over the years. Some examples are: \cite{johansson_interactive_1998,guzman_interactive_2008,esteban-escano_use_2022,paharia_interactive_2022,rosas_pedagogical_2022}, but many more exist. The recent book by \cite{guzman_automatic_2023} explains one of the most advanced and diverse interactive tools aimed at teaching. LU-PZE, presented in this paper, differs from these other tools because it is run in the web browser, making it notably easier to start using than interactive tools that require software installation. 
LU-PZE, as well as many of the above-mentioned tools, visualize multiple concepts from the course simultaneously.
Thus, using LU-PZE, the students work actively with the course content and learn to see connections between different concepts, thereby gaining a deeper knowledge of the course.
The control theory concepts that are visualized and that the students can explore in LU-PZE are \citep{glad_control_2000}:

\begin{itemize}
    \item Transfer functions
    \item Poles and zeros
    \item Step/impulse responses
    \item Bode plots
    \item Nyquist diagram
\end{itemize}

The LU-PZE interface and features are explained in Section~\ref{sec:lupze}. The gamified elements of LU-PZE are explained in Section~\ref{sec:education}. Details on the implementation and mathematical approach are explained in Section~\ref{sec:method}. The results are presented in Section~\ref{sec:results}, the discussion around LU-PZE in Section~\ref{sec:discussion}, and some concluding words in Section~\ref{sec:conclusion}.

\section{LU-PZE interface}\label{sec:lupze}
An overview of the default screen in the Lund University Pole-Zero Explorer (LU-PZE) tool is shown in Figure~\ref{fig:whole_screen}. The two plots to the left are the Bode plots, with the magnitude plot at the top and the phase plot at the bottom. The top plot in the middle is the step response, and at the bottom is the Nyquist diagram. The pole-zero plots for the systems are to the right of the step response plot. To the far right is a panel with transfer functions and sliders to change the parameters of the transfer functions. At the top of the right panel is a dropdown menu to change the input to the system.
At the bottom of the screen, you'll find the gain margin and phase margin values for the selected transfer function. Changes made to a system's parameters in one plot are instantly reflected in the other plots, ensuring a comprehensive and consistent view of the system. In the menu bar at the top right are icons to go to the quiz, assignments, achievements, fullscreen, add more transfer functions, and so on. 

The following subsections describe the features within each part of LU-PZE.

\subsection{Transfer functions}
The transfer functions shown in the default view are
\begin{align*}
    G_1(s) &= \frac{k_1}{1+T_1s}\\
    G_2(s) &= \frac{k_2}{(1+T_2s)(1+T_3s)}\\  
    G_3(s) &= \frac{k_3\omega_0^2}{s^2+2\zeta\omega_0s+\omega_0^2}\\
    G_4(s) &= \frac{3}{1+s}e^{-Ls}
\end{align*}
where $G_1$ describes a first-order system, $G_2$ a second-order system with real poles, $G_3$ a second-order system with complex poles, and $G_4$ a first-order system with time delay.

The following additional pre-defined transfer functions are available by pressing the plus \includegraphics[height=\myheight]{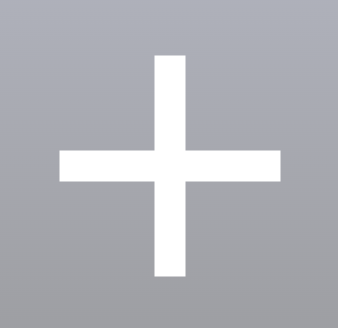} sign at the top of the right panel
\begin{align*}
    G_5(s) &= \frac{k_4(1+T_8s)}{(1+T_6s)(1+T_7s)}\\
    G_6(s) &= \frac{1}{(1+T_5s)^4}
\end{align*}
where $G_5$ describes a system with a zero and two real poles, and $G_6$ a system with four identical poles.

The parameter $k_i$ in the transfer functions is the gain of the corresponding system and can be adjusted with a slider for each system. The parameter $T_i$ is the corresponding system's time constant(s) and can be adjusted with a slider for each system. The parameter $\omega_0$ is the self-oscillatory frequency, and $\zeta$ is the relative damping of the second-order system and can be adjusted with sliders for the second-order system with complex poles. The pre-defined transfer functions are those that are taught in most courses on automatic control. In addition to these, the student can write their own transfer functions or edit the pre-defined ones.

Pressing the export icon \includegraphics[height=\myheight]{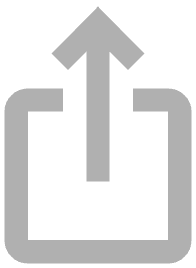} beside the closing cross \includegraphics[height=\myheight]{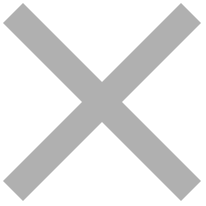} for each transfer function opens a window with Python, MATLAB, or Julia code that can be downloaded and run elsewhere. 

\subsection{Pole-zero plots}
To the left of the panel with transfer functions are the pole-zero maps. Crosses represent the poles and circles zeros. The pole and zero markers can be dragged to change the location of the pole/zero. 

\subsection{Step/impulse response}
The input to the system is, by default, a step response but can be changed to an impulse response from the drop-down menu in the top right panel. Here, the default step response view is described, which is shown at the top center of the screen. The thick lines show the step response of the systems, the crosses represent the time constants for the systems, and the thin horizontal lines show the static gain of the systems. 

When you hover over the plot, a vertical line appears, indicating the output of the response at that specific point in time. The hovering point snaps to the step response curve if you hover close to a step response. A dot to the right represents the static gain for a system with a corresponding dot in the Bode magnitude plot. 

For $G_1$ and $G_2$, the systems' gain and time constants can be changed by dragging the plots. For $G_3$, the $\omega_0$ parameter can be adjusted by dragging the step response to the right and left, and the $\zeta$ parameter can be adjusted by dragging the step response up and down. For $G_4$ the time delay is changed by dragging the step response to the right and left. 

\subsection{Bode plot}
The Bode plots are shown to the left of the screen, with the magnitude curve at the top and the phase curve at the bottom. The crosses correspond to the systems' poles and can be dragged around to change the parameters of the systems. 

When hovering over the plots, the frequency, as a vertical line, and the magnitude/phase, as a horizontal line, for that coordinate is shown. Dots appear in the Nyquist diagram representing the frequency at the Nyquist diagrams. When you hover over the Bode magnitude plot, a circle appears in the Nyquist diagram, illustrating the amplitude of your hovering. If you hover over the Bode phase plot, an arc appears in the Nyquist diagram, illustrating the phase of your hovering. If you follow one of the systems' Bode plots, the corresponding circle/arc is shown in the Nyquist diagram for that system.

For $G_1$ and $G_2$, the systems' gain can be changed by dragging their magnitude plot up and down and the time constants by dragging the crosses. For $G_3$, the magnitude can't be changed; only the cross can be dragged, corresponding to changing $\omega_0$ and $\zeta$. For $G_4$, no alterations except the time delay are provided, and adjusting the Bode plots has no effect.

\subsection{Nyquist diagram}
The Nyquist diagram is shown below the step response at the bottom center of the screen. The crosses represent the poles of the system but cannot be dragged to change the poles' value. The critical point -1 for the Nyquist criteria is also shown in the plot. 

When hovering over the Nyquist plot, a line and an arc appears, representing that coordinate's phase and amplitude. In the Bode plots, vertical lines appear (one in the magnitude plot and one in the phase plot), representing the same amplitude/phase as hovered over in the Nyquist plot.

\subsection{Stability information}
The amplitude and phase margins, written with their corresponding crossover frequencies, are at the bottom of the screen, below all plots. Press the system name to change which system to show.

\section{Gamifying control education} \label{sec:education}
Gamification of education means that the course, or parts thereof, is made into a system where the students get points for reaching specified goals. The students could get points for completing a task or learning a new skill. As mentioned in the introduction, gamification of education can help increase student motivation and learning \citep{zeybek_gamification_2024,buckley_gamification_2016}. Built into LU-PZE are different gamification aspects, explained in the following sections. There are achievements the student can get, assignments the student can complete, and quiz questions the student can do. When completing these challenges, the student gets points and reaches a higher level, eventually gaining the gold badge in all categories. As an aid for student learning, right-clicking opens the questions and hints functionality of LU-PZE.

\subsection{Encouraging exploration: Achievements}
The student gets stars, \includegraphics[height=\myheight]{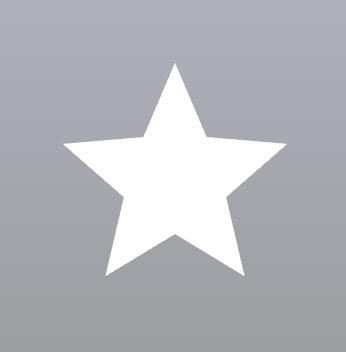}, for exploring the LU-PZE tool, for example, by dragging the poles in a pole-zero map, hovering in the Nyquist diagram, or adding a new transfer function. The purpose of these achievements is to encourage the student to explore the tool and give some hints for what can be done. For example, the students get an achievement for downloading the Python code for a transfer function, a functionality in the LU-PZE tool that is not apparent at first glance. 

\subsection{Traditional in-order exercises: Assignments}
Under the assignments menu, \includegraphics[height=\myheight]{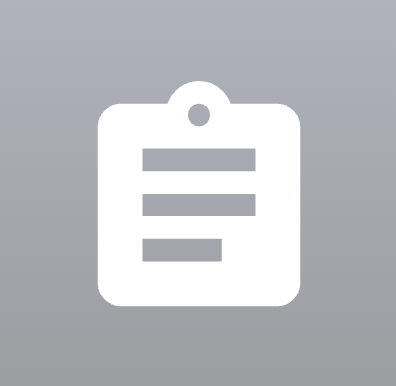}, there are assignments related to the different kinds of systems, $G_i$, illustrated in the LU-PZE tool. The student can complete the assignments in any order, but the assignments for the first-order system are easier than the rest, and the idea is that the students should start with these first-order system assignments. The tasks within one assignment group are presented in order of increasing complexity, and they are loosely built on knowledge from earlier assignments. Each task, when solved, presents the student with motivational text on why this task is important to master and some explanation for the solution. Below are two examples of tasks for the first-order system:

\begin{itemize}
\item Your task is to change T1 by moving the slider or type in T1's value to make the pole's location -1/2 in the pole-zero map. Can you explain how the pole's location and its time constant are related?
\item Drag the pole marker in the pole-zero map to make the system four times faster than the orange one. What does it mean to have a faster system?
\end{itemize}

The assignments require the students to change the system in one domain to observe the system in another, e.g., by dragging the bode plot to reach a desired step response. This type of assignment aids the students in discovering correlations between the different representations of a system, which helps them gain a deeper knowledge of control theory.

\subsection{Volume training: Adaptive Quiz}
The quiz, \includegraphics[height=\myheight]{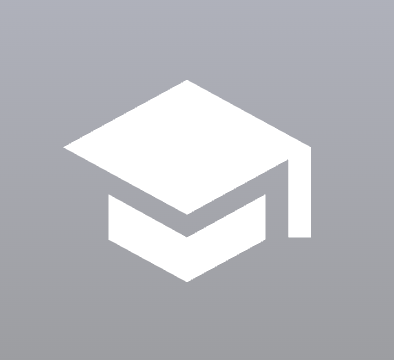}, is aimed at providing students with a level of intensive volume training that is difficult to obtain through alternative methods. 
At the time of this paper's writing, we have implemented these main categories of quiz questions:
\begin{itemize}
    \item Click on a certain frequency
    \item Click on a certain time
    \item Click on a certain Nyquist angle
    \item Click on a system with a certain complexity
    \item Click the odd-one-out
\end{itemize}

The quiz tasks are presented one at a time, with a difficulty level that adapts to the student's prior performance. When a correct answer is given, the difficulty level is increased for that section based on the streak of correct answers within that section. A wrong answer lowers the difficulty and resets the streak counter. This way, a student is allowed to have separate difficulty levels in different sections. A correct answer is immediately awarded with the next question. A wrong answer will present the student with text explaining what the student answered and why this answer does not answer the question. The explaining text is a highly dynamic response that is tailored to this specific combination of question and answer. 
The adaptive difficulty is to keep the student immersed by providing just enough challenge for each category of quiz questions. In the case of a group seminar or exercise with a teacher, a student's results are shown on screen to help the exercise leader understand the students' performance.

\subsection{Questions}
In the LU-PZE tool, right-clicking opens the questions and hints functionality of the tool. The purpose of this is for the students to easily have access to theoretical explanations of the concepts they observe in LU-PZE. Once right-clicked, the screen is filled with question marks, \includegraphics[height=\myheight]{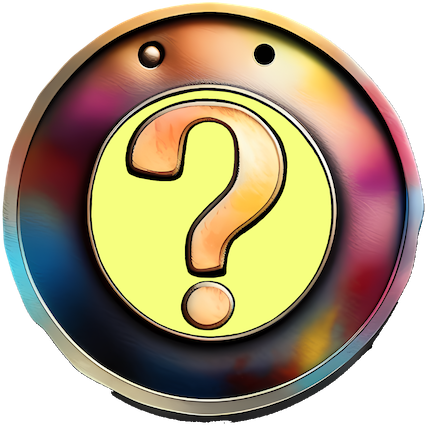}, which indicate locations of questions. The students can view the questions along with their corresponding answers by hovering over the question marks. The answers are presented in three levels of detail: a quick summary overview, an expanded explanation with practical examples, and a mathematically-oriented response complete with equations, formulae, and theoretical proofs. This allows students the flexibility to choose whether or not they find the question relevant and determine how much time they wish to spend reading the answer.

\section{Method}\label{sec:method}

\subsection{Mathematical approach}
LU-PZE is implemented with the transfer function equations as the foundation. From these, the Bode magnitude and phase plots are mathematically relatively easy to calculate by substituting the complex frequency domain variable $s$, found in the equations, with $j\omega$, and calculating for $\omega = [10^{-2} \cdots 10^{3}]$. The resulting complex value has an amplitude and a phase that can be plotted. For the phase plot, the complex value will only give us phases in the range $-\pi \leq \phi < \pi$, and we will need to track the phase changes and add or subtract a multiple of $2\pi$ when needed, which can be troublesome with steep phase graphs, typically found in time delayed transfer functions like $\frac{1}{1+s}e^{-Ls}$ at higher frequencies and large time delays.

The Nyquist diagram is also fairly fast and easy to calculate, as it is based purely on the Bode magnitude and phase values. The idealized Nyquist diagram should contain the frequencies $\omega=0~rad/s$ and $\omega=\infty$~rad/s as well. However, for general transfer functions, computers are not well equipped to handle equations with infinite values, and the last pixels at the edges of graphs in the Nyquist plot can fairly easily be extrapolated by the student.

Calculating the step response for any freely given transfer function is a harder challenge. We need to calculate the inverse Laplace transform, which has a known solution for a few well-studied lower degree polynomial transfer functions \citep{cohen_2007}. LU-PZE uses the Gaver-Stehfest algorithm for approximating the time response \citep{stehfest_1970}. This algorithm uses abscissa along the real axis of the complex plane to estimate the time-domain function. This method does not work well with oscillatory functions as they have poles located closer to the imaginary axis. The Gaver-Stehfest method also depends on the summation of terms in a series that grows very large, leading to numerical cancellation. By calculating ten transfer function values along the real axis, one single time response value can be approximated. This is compute intensive, even for a laptop, and to keep the graphs' updates smooth while dragging graphs and sliders in the GUI, the transfer functions with known step responses, $G_1(s)$ to $G_4(s)$, are calculated using known time response formulas. For instance, the first-order system with one pole
\begin{align*}
    G_1(s) = \frac{k_1}{1+T_1s}
\end{align*}
will have the time step response $$g_1(t) = k_1 (1 - e^{-\frac{t}{T_1}}).$$
This formula is analytically accurate and approximately 20 times faster than using the Gaver-Stehfest approximation method, which may yield inaccuracies.

\subsection{Implementation}
LU-PZE is implemented in JavaScript, CSS and HTML. A number of JavaScript libraries have been used, mainly p5.js\footnote{\url{https://p5js.org/}} for graphics, Math.js\footnote{\url{https://mathjs.org/}} for computations, and MathLive\footnote{\url{https://cortexjs.io/mathlive}} for visualizing LaTeX equations. The LU-PZE JavaScript code was built upon the now defunct Bowde\footnote{\url{https://github.com/nnnawi/BOWDE}}, and the GUI elements and handling was inspired by Desmos\footnote{\url{https://www.desmos.com/calculator}}.

The main implementation challenge has been making the graphical output as graphics-card-friendly as possible to make the rendering close to real-time and. In order to make the source code as approachable as possible, we have chosen to keep the development environment as close to vanilla JavaScript as possible, devoid from frameworks such as jQuery, React, Angular, Vue, and Svelte, and free from build systems and package managers such as NPM, Grunt, Gulp, Yarn or webpack. No additional downloads or installations are required to develop LU-PZE further. LU-PZE's \texttt{index.html} can be run as a file in the browser with full functionality.

\subsection{Source Code}
The source code is available at \\ \url{https://github.com/lu-pze/lu-pze.github.io}

\section{Results}\label{sec:results}
LU-PZE is used in all basic automatic control courses held at Lund University\footnote{\url{https://kurser.lth.se/lot/course/FRTF05}}, taken by roughly 1000 students per year, and can be freely accessed at \\ \url{https://lu-pze.github.io}

Another paper will examine the students' satisfaction with LU-PZE and their (hopefully) improved understanding of automatic control theory as a result of using LU-PZE.

\section{Discussion}\label{sec:discussion}
The main difference between LU-PZE and other existing interactive control tools, such as the one described in the book Automatic Control with Interactive Tools by \cite{guzman_automatic_2023}, is that LU-PZE is run in a web browser and does not need any installation, making the system extraordinarily easy for students to use. The purpose of LU-PZE is to be an easy-to-use tool that gives the students a visual aid for fundamental concepts of automatic control, adapting to the student's knowledge level. Some elements of LU-PZE are gamified to increase the students' motivation to explore the tool and, thereby, the automatic control theory. If the students need more advanced features for analysis, applications such as the one described in the above-mentioned book Automatic Control with Interactive Tools by \cite{guzman_automatic_2023} is recommended, or coding the problem in Python/MATLAB/Julia or any other preferred programming language.

\subsection{Comparision to other tools for active learning and online assessment}
LU-PZE serves as a study aid that promotes active learning and exploration. \emph{Threshold assessment} determines whether students have acquired the vital knowledge, skills, and abilities required for successful course completion, and will need to be performed through other methods than LU-PZE. Computerized formative assessments, whether conducted online or offline, can induce stress and occasionally lead to cheating. We have decided to omit assessment in LU-PZE, allowing students to use the tool anonymously without being monitored by the system or their supervisors. Computer aided assessment systems with written questions are easily 'bypassed' using modern AI tools such as chatGPT. We suggest using LU-PZE either as a self-paced tool for the students to use freely at home, or in a classroom environment with a teacher 
guiding discussions within student groups. Furthermore, text-based AI tools like chatGPT offer limited assistance in understanding and solving visually intensive concepts of automatic control theory, such as gain margin, phase margin, Nyquist diagrams, and Bode plots, which involve manipulating time constants, gains, pole and zero locations.

In a classroom teaching session, \emph{Blackboard quizzes}, \emph{Mentimeter}, \emph{TurningPoint}, and \emph{Kahoot} exemplify  classroom response systems (CRS), transforming a pre-made set of questions into electronically collected, shared discussion material. Such tools are invaluable for breaking the monotony of lectures and initiating discussions. In this context, LU-PZE is better suited as an academic \emph{show-and-tell}, or scholarly symposium, where the teacher can introduce concepts of automatic control theory by discussing and solving LU-PZE assignments in front of the class. As LU-PZE is a simple web page, integrating it into a presentation session is an effective way to engage students and motivate them to explore automatic control concepts independently.

\subsection{Limitations}
LU-PZE needs a specific amount of screen space and works best on a computer screen in landscape mode. On older computers with low-resolution screens, certain menu elements may be hidden using the fullscreen button \includegraphics[height=\myheight]{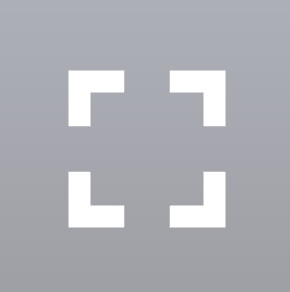} among the top-right icons. Devices such as iPads and Android tablets do not support hovering over graphs due to the absence of a mouse pointer, although some tablets with stylus support may offer this feature. Handheld devices generally do not have keyboards, which limits their capacity to edit and input transfer functions. To enhance the frame rate and smoothness of the graphical user interface (GUI) on older computers, it may be necessary to hide or remove graphs.

As mentioned previously, time response graphs of general transfer functions entered manually are approximated using the Gaver-Stehfest algorithm. For a more precise analysis, programs languages such as Python, MATLAB, or Julia should be used.

Using the exported source code for plotting will require additional downloads and installation. Python and Julia are free to install and use, but depending on your level of computer proficiency, they can be challenging to install and run. Using MATLAB without access to the often generous university licensing can be excessively expensive.

LU-PZE is a playground for understanding transfer functions of open-loop systems. For analyzing closed loop systems with feedback control, other tools should be used.

\section{Conclusion}\label{sec:conclusion}
LU-PZE is an interactive tool for illustrating fundamental concepts related to control theory. It covers the relation between transfer functions, pole-zero plots, step responses, Bode plots, and Nyquist diagrams. The tool is straightforward to use since it is web-based. The LU-PZE tool can be run at: \url{https://lu-pze.github.io}

\section*{Contributions}
\textbf{Pex Tufvesson}: Conceptualization, Methodology, Software, Writing - Review \& Editing, and Visualization.\\ \textbf{Frida Heskebeck}: Conceptualization, Writing - Original Draft, and Writing - Review \& Editing.

\begin{ack}
The authors want to thank the people at the Department of Automatic Control at Lund University for providing feedback. 
\end{ack}

\bibliography{ifacconf}             
\end{document}